\documentclass[11pt]{article}
\usepackage{hyperref}
\pdfoutput=1
\begin{document}
\title{UFO: ``Unidentified'' Floating Object Driven by Thermocapillarity}
\author{Yuejun Zhao and Chuan-Hua Chen \\
\\\vspace{6pt} Department of Mechanical Engineering and Materials Science,
\\ Duke University, Durham, NC 27708, USA}
\maketitle
\begin{abstract}
In this fluid dynamics video, we show thermocapillary actuation of a binary drop of water and heptanol where the binary drop in motion takes on a UFO-like shape. On a parylene-coated silicon surface subjected to a linear temperature gradient, a pure heptanol droplet quickly moves to the cold side by the Marangoni stress, while a pure water droplet remains stuck due to a large contact angle hysteresis. When the water droplet was encapsulated by a thin layer of heptanol and thermally actuated, the binary droplet takes on a peculiar shape resembling an UFO, i.e. an ``unidentified'' floating object as the mechanism is not yet completely understood. Our finding suggests that pure liquid droplets (e.g. aqueous solutions) that are not conducive to thermocapillary actuation can be made so by encapsulating them with another judiciously chosen liquid (e.g. heptanol).
\end{abstract}
\section{Introduction}
\href{http://ecommons.library.cornell.edu/bitstream/1813/8237/2/LIFTED_H2_EMS
T_FUEL.mpg}{Video
1} includes all necessary information and videos.

\end{document}